\documentclass[10pt,conference]{IEEEtran}
\IEEEoverridecommandlockouts
\usepackage{amsmath,amssymb,amsfonts}
\usepackage{url}
\usepackage{tabularx}
\usepackage{enumitem}
\usepackage[skip=0pt]{caption}
\usepackage{setspace}
\usepackage{multirow}
\usepackage{amssymb}
\usepackage{booktabs}
\usepackage{bbm}
\RequirePackage{subcaption}

\usepackage{algorithmicx}
\usepackage[ruled,vlined,linesnumbered]{algorithm2e}
\algnewcommand\algorithmicinput{\textbf{Input:}}
\algnewcommand\algorithmicoutput{\textbf{Output:}}

\usepackage{graphicx}
\usepackage{textcomp}
\usepackage{xcolor}

\usepackage[round,sort]{natbib}
\usepackage{color}

\def\BibTeX{{\rm B\kern-.05em{\sc i\kern-.025em b}\kern-.08em
    T\kern-.1667em\lower.7ex\hbox{E}\kern-.125emX}}
\begin{document}

\title{Audit-LLM: Multi-Agent Collaboration for Log-based Insider Threat Detection}

% \author{\IEEEauthorblockN{Anonymous Authors}}

\author{\IEEEauthorblockN{Chengyu Song}
\IEEEauthorblockA{\textit{Systems Engineering Institute} \\
\textit{Academy of Military Sciences}\\
Beijing, China \\
songchengyu@alumni.nudt.edu.cn}
\and
\IEEEauthorblockN{Linru Ma*}
\IEEEauthorblockA{\textit{Systems Engineering Institute} \\
\textit{Academy of Military Sciences}\\
Beijing, China \\
malinru@163.com}
\and
\IEEEauthorblockN{Jianming Zheng$^{\star}$}
\IEEEauthorblockA{\textit{State Key Laboratory of } \\
\textit{Mathematical Engineering}\\
\textit{and Advanced Computing}\\
Wuxi, China \\
zhengjianming12@nudt.edu.cn}
\and
\IEEEauthorblockN{Jinzhi Liao}
\IEEEauthorblockA{\textit{National University of Defense Technology} \\
\textit{National Key Laboratory of}\\
\textit{Information Systems Engineering}\\
Changsha, China \\
liaojinzhi12@nudt.edu.cn}
\and
\IEEEauthorblockN{Hongyu Kuang}
\IEEEauthorblockA{\textit{Systems Engineering Institute} \\
\textit{Academy of Military Sciences}\\
Beijing, China \\
khy\_y@qq.com}
\and
\IEEEauthorblockN{Lin Yang}
\IEEEauthorblockA{\textit{Systems Engineering Institute} \\
\textit{Academy of Military Sciences}\\
Beijing, China \\
yanglin61s@126.com}
}

\maketitle

\begin{abstract}
Log-based insider threat detection (ITD) detects malicious user activities by auditing log entries. 
Recently, Large Language Models (LLMs) with strong common sense knowledge are emerging in the domain of ITD.
Nevertheless, diverse activity types and overlong log files pose a significant challenge for LLMs to directly discern malicious ones within myriads of normal activities. 
Furthermore, the faithfulness hallucination issue from LLMs aggravates its application difficulty in ITD, as the generated conclusion may not align with user commands and activity context.  
In response to these challenges, we introduce Audit-LLM, a multi-agent log-based insider threat detection framework comprising three collaborative agents:
(i) the Decomposer agent, breaking down the complex ITD task into manageable sub-tasks using Chain-of-Thought (COT) reasoning;
(ii) the Tool Builder agent, creating reusable tools for sub-tasks to overcome context length limitations in LLMs;
and (iii) the Executor agent, generating the final detection conclusion by invoking constructed tools.
To enhance conclusion accuracy, we propose a pair-wise Evidence-based Multi-agent Debate (EMAD) mechanism, where two independent Executors iteratively refine their conclusions through reasoning exchange to reach a consensus.
Comprehensive experiments conducted on three publicly available ITD datasets—CERT r4.2, CERT r5.2, and PicoDomain—demonstrate the superiority of our method over existing baselines and show that the proposed EMAD significantly improves the faithfulness of explanations generated by LLMs.
\footnote{* Corresponding author}
\footnote{$^{\star}$ Equal contribution}
\end{abstract}


%!TEX root = ./Audit-LLM.tex
\section{Introduction}
\label{Introduction}

Insider threats are one of the most challenging attack patterns in practice as they are usually carried out by authorized users who have legitimate access to sensitive and confidential materials~\citep{DBLP:journals/csur/HomoliakTGEO19}.
To address the task, Insider Threat Detection (ITD) is coined to detect malicious activities by insiders, involving monitoring and analyzing logs. 
These logs contain critical records of various user behaviors essential for troubleshooting and security analysis.

% Conventional ITD models are not ready in this light~\citep{DBLP:conf/icse/0001TMYZY24}.
%
% Conventional ITD models, e.g., DeepLog~\citep{DBLP:conf/ccs/Du0ZS17}, rely on Deep Learning(DL)-based representation learning algorithms to capture the multi-dimensional behavioral characteristics of users~\citep{DBLP:journals/compsec/YuanW21}.
% %
% However, they encounter two major challenges: \textbf{overfitting} and \textbf{opacity}.
% %
% Regarding overfitting, malicious activity from insiders is sporadic compared to benign activity in real-world scenarios~\citep{DBLP:journals/compsec/YuanW21}. 
% %
% As a result, deep learning-based models often become biased towards learning and predicting the majority benign class, which dominates the training data. 
% %
% This bias can lead to inadequate detection of rare but critical malicious behaviors.
% %
% Concerning opacity, these models usually deliver results in a black box format, which is insufficient for practical log auditing.  
% %
% However, effective log auditing requires not only the detection of malicious behavior but also interpretable evidence that enhances the credibility and actionable insight of the results. 
% %
% The lack of interpretability diminishes the credibility of the model's outputs, consequently impairing the efficiency of security auditing efforts.

Conventional ITD models
% like DeepLog
\citep{DBLP:conf/ccs/Du0ZS17,DBLP:conf/sp/LeZH21,DBLP:journals/tifs/LiLJLYGY23} utilize Deep Learning for capturing diverse user behavioral characteristics~\citep{DBLP:journals/compsec/YuanW21}. 
However, the inherent problems in this line of approaches, i.e., overfitting and opacity, hinder the further enhancement of their performance.
% they face challenges with overfitting and opacity.
The emergence of overfitting is caused by the scarcity of insider threats 
% arises because insider threats are infrequent
in comparison to benign activities, resulting in a bias towards benign behavior while neglecting 
% This bias can overlook 
critical malicious activities. 
Opacity limits practical log auditing by delivering results in an opaque format, lacking the interpretability necessary for credible and actionable insights in security auditing.

In response, there is a booming trend of applying LLMs in the domain of ITD~\citep{DBLP:journals/corr/abs-2306-01590,DBLP:conf/hpcc/QiHLYFYQSXW23,DBLP:journals/corr/abs-2403-00878}.
Leveraging LLMs' extensive commonsense knowledge and capacity for intricate multi-step reasoning, existing methods require them to either justify each decision, thereby implicitly constructing logical chains of reasoning \citep{DBLP:conf/icse/0001TMYZY24}, manually define intermediate steps for log auditing \citep{DBLP:conf/hpcc/QiHLYFYQSXW23}, or use a few annotated log samples to provide context and guide predictions \citep{DBLP:conf/icse/0001TMYZY24}.
% LLMs~\citep{DBLP:conf/nips/Ouyang0JAWMZASR22,DBLP:journals/corr/abs-2302-13971} naturally possess rich commonsense knowledge and capable of sophisticated multi-step reasoning. 
These abilities empower them to conduct log auditing in a zero-shot manner without the need for training or fine-tuning, thereby fundamentally mitigating the risks of overfitting caused by highly imbalanced categories. 
Additionally, auditing results can be delivered in a pre-defined human-readable format, thereby avoiding issues of non-interpretable outcomes.

Despite the promising applications of LLMs in ITD, current studies merely transfer them in a straightforward input-output way.
This direct transformation fails to approach the specific features of ITD.
Specifically, we identify challenges as follows:
(1) Malicious behaviors exhibited by users are \textbf{intrinsically manifold}. Insiders may leak confidential data through the network, mobile storage devices, or email~\citep{DBLP:conf/sp/GlasserL13}, requiring a multidimensional examination of user behaviors. 
% there remains a critical need to address certain overlooked challenges.
% First, albeit a binary classification in appearance, the ITD task is notoriously challenging for direct analysis using LLM.
% \scy{Existing LLM-based methods~\citep{DBLP:conf/hpcc/QiHLYFYQSXW23,DBLP:conf/kbse/LeZ23} fail to} conduct  may result in overlooking these incidents. 
(2) The input length constrained by LLMs might result in \textbf{inadequate detection}. 
% Considering the limitations in computational resources and response speed, 
Online APIs for LLMs commonly restrict the size of input windows (e.g., approximately 128K tokens for GPT-4), which proves inadequate for incorporating entire overlong logs, thereby disregarding crucial contextual details such as typical user behaviors. 
% However, truncating log lengths \scy{by current methods~\citep{DBLP:conf/icse/0001TMYZY24}} to fit these input constraints results in \textbf{inadequate detection} of crucial contextual details, such as typical user behaviors.
% This limitation impedes the model's ability to distinguish between normal and abnormal behaviors effectively.
(3) The \textbf{faithfulness hallucination} brought by LLMs leads to
% \scy{Last but not least, LLM-based methods} face \textbf{faithfulness hallucination} issues~\citep{DBLP:journals/corr/abs-2311-05232}, referring to 
the divergence of generated content from user instructions.
%For example, the sub-task's results clearly are identified as malicious behavior by the user, yet the agent concluded: ``benign.''
%
For example, even if some sub-tasks' results are identified as malicious, there remains a possibility for LLMs to categorize the whole log set as benign.

\begin{figure}
	\centering
	\includegraphics[width=0.48\textwidth]{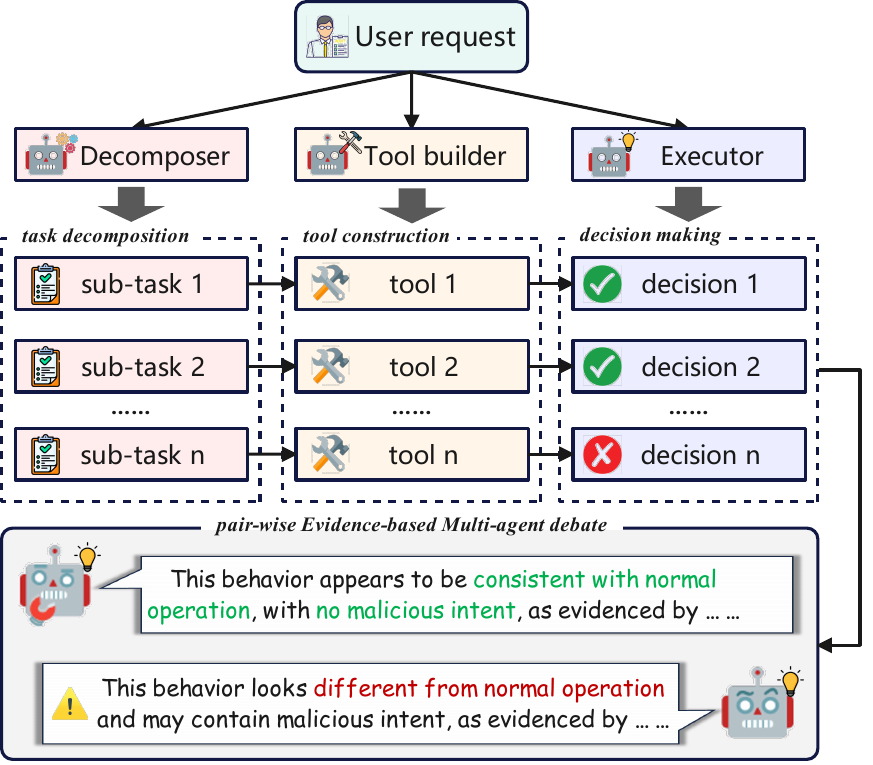}
	\caption{An example of the three agents included in Audit-LLM, along with their interaction and workflow.}
	\label{introduction}
\end{figure}

Considering the multiple perspectives involved in identified challenges, distinct capabilities are required to collaboratively handle ITD, i.e., decomposing log auditing into sub-tasks, accessing information beyond the input window, and possessing mechanisms to mitigate hallucinations. 
The requirement is in line with the main idea of multi-agent systems~\citep{DBLP:conf/aaai/0009SFTLL024,DBLP:conf/acmturc/Deng0DW24}, where several specialized agents are played as specific roles to achieve a shared goal collectively.
Thus, we draw upon the workflows of human log auditors to develop a multi-agent insider threat detection framework \textbf{Audit-LLM}.
Enhanced by multi-agent collaboration, the Audit-LLM framework focuses on task decomposition, tool creation, and hallucination elimination, as depicted in Fig.~\ref{introduction}.

% To overcome these obstacles, the model must decompose ITD into sub-tasks using CoT for comprehensive behavioral detection, access information beyond the input window to address log truncation, and possess mechanisms to mitigate hallucinations.

Specifically, instead of achieving the final result in a single step, we instruct a Chain-of-Thought~\citep{DBLP:conf/aaai/JiLDN24} (CoT)-oriented agent, referred to as the \textbf{Decomposer}, to tailor the complex ITD task into a series of sub-tasks.
%according to available types of log entries. 
% By coincidence, this strikes a chord with the current trend of building a Chain-of-Thought (CoT) ~\citep{DBLP:conf/aaai/JiLDN24} process for complex tasks. 
This agent facilitates a comprehensive evaluation of user behavior from multiple perspectives.
%Second, 
% like LLMs constrained by window limits, 
% humans have a finite capacity to process information simultaneously. 
% The limitation is managed by employing various tools, such as intrusion-detection systems, to extract, analyze, and synthesize valuable information from the log entries.
Expanding on this analogy, we guide the next agent, known as the \textbf{Tool Builder}, to develop a suite of sub-task-specific tools to extract global characteristics for detection.
These tools are engineered to derive insights from the log set, such as a user's historical login frequency and the verification of website legitimacy, thereby improving the final conclusion.
Lastly, we develop a third agent named \textbf{Executor}, tasked with systematically accomplishing sub-tasks by invoking constructed tools to realize threat detection.
%dynamically invoking tools, systematically executing predefined sub-tasks, and reaching the final conclusion.
Drawing inspiration from the human ``peer review'' process, which improves the quality and reliability of work through mutual evaluations and feedback, we propose a pair-wise Evidence-Based Multi-Agent Debate (EMAD) mechanism to mitigate the faithfulness hallucination issue encountered in LLMs employed for ITD.
% Pair-wise EMAD involves two independent Executors to conduct debate by iterative feedback, facilitating refinement and ensuring the generation of accurate explanations, guiding conclusions toward genuine rationales.

In this paper, we introduce Audit-LLM (a multi-agent log-based insider threat detection framework) that integrates the aforementioned ideas. Our contributions are three-fold:

\begin{itemize}
	\item To the best of our knowledge, we are the first to employ multi-agent collaboration for ITD and propose Audit-LLM, a multi-agent log-based insider threat detection framework.
	\item We counter faithfulness hallucination issues by introducing a pair-wise Evidence-based Multi-agent Debate mechanism. This enables agents to engage in an iterative refining process, thus bolstering the reliability of our ITD system.
	\item We evaluate the proposed Audit-LLM alongside state-of-the-art baselines for the ITD task using three publicly accessible datasets. Our findings demonstrate the superiority of Audit-LLM compared to the competitive baselines.
\end{itemize}

\section{Related Work}
\label{relatedwork}

In this section, we first review related works of traditional ITD methods and deep learning-based ITD methods in Sec.~\ref{ITD}.
Then, in Sec.~\ref{L4C}, we provide a detailed discussion of various approaches for applying LLMs in the field of cybersecurity.

\subsection{Log-based insider threat detection}
\label{ITD}
Insider threat detection has attracted considerable research interest over the last decade as an important task in cybersecurity.
Over the years, extensive research has been conducted to develop effective approaches for detecting insider threats. Broadly, these approaches can be categorized into two main streams: traditional methods and deep learning methods.

Traditional ITD methods can further be classified into two types: anomaly-based and misuse-based approaches.
Anomaly-based detection is the prevalent approach.
For instance, 
\citet{DBLP:conf/sp/BrdiczkaLPSPCBD12} propose a traitor assessment using Bayesian techniques that combined structural anomaly detection from information and social networks with psychological profiling.
Additionally, \citet{DBLP:journals/tifs/CaminaMTM16} propose detection systems for masqueraders utilizing SVM and KNN as one-class techniques. 
In contrast, misuse-based methods incorporate softer forms of matching through similarity measurement. 
For instance, 
\citet{DBLP:conf/ccs/AgrafiotisEGC16} propose tripwire grammar capable of capturing abstraction of policies that organizations adopted, as well as signatures of insider misbehaviors.
Moreover, \citet{DBLP:conf/inc/MagklarasF12} design an insider threat prediction and specification language (ITPSL), which has markup features and utilizes logical operators.

With the development of deep learning methods, there is a departure from traditional approaches, as practitioners increasingly turn to neural networks to distinguish between benign and malicious behaviors.
For instance, \citet{DBLP:conf/bigdataconf/YuanZWL19} employ hierarchical neural temporal point processes to capture activity types and time information within user sessions. 
Further, \citet{DBLP:conf/ccs/LiuWZJXM19} introduce a network security threat detection method based on heterogeneous graph embedding. It achieves user behavior detection by constructing a heterogeneous graph, conducting graph embedding learning, and employing detection algorithms.
Moreover, \citet{DBLP:journals/ijon/FangWFH22} present LMTracker for lateral movement path detection based on heterogeneous graphs and propose a representation method for lateral movement paths and devise an unsupervised detection algorithm utilizing reconstruction error.

% Traditional methods struggle with high-dimensional and complex user behavior data, limiting their effectiveness in identifying insider threats. 
% %
% In contrast, deep learning methods require large amounts of labeled data for training, which is impractical due to the rarity of malicious insider behavior. 
% %
% Instead, Audit-LLM leverages the extensive knowledge and zero-shot generation capabilities of LLMs to accurately detect anomalous behaviors in log data without needing pre-training.

However, in real-world scenarios, malicious insider threat behaviors are extremely rare compared to benign behaviors.
This rarity can lead deep learning models to exhibit bias, often favouring predictions of benign activity.
Furthermore, deep learning typically outputs log classification results in an end-to-end manner, lacking interpretable intermediate results that are crucial for end-users such as auditors to trust.
Instead, Audit-LLM leverages the extensive knowledge and zero-shot generation abilities of LLMs to accurately detect malicious behaviors and provide an interpretable analysis process.

\subsection{LLM for cybersecurity}
\label{L4C}

The constantly changing landscape of modern cybersecurity poses significant challenges, with adversaries adapting tactics to exploit vulnerabilities and avoid detection.
However, AI advancements, especially Large Language Models (LLMs), offer promising avenues for strengthening cybersecurity, serving not only as defensive measures but also as offensive tools.
For instance,
\citet{DBLP:journals/corr/abs-2403-01038} present a system named AutoAttacker, which leverages Large Language Models for automated network attacks, utilizing language models for planning, summarization, navigation, and experience management. The paper proposes a system for automated penetration testing.
Moreover, \citet{DBLP:journals/corr/abs-2404-08144} investigate the capability of LLMs to automatically exploit cybersecurity vulnerabilities. 
Employing the GPT-4 model in conjunction with CVE vulnerability descriptions, they were able to successfully exploit 87\% of real-world software vulnerabilities.
Concurrently, LLMs can serve as potent instruments for the defensive side, aiding in the detection and identification of potential security threats.
For instance,
\citet{DBLP:journals/corr/abs-2403-00878} leverage large language models to enhance strategic reasoning capabilities in cybersecurity, realizing a comprehensive human-machine interactive data synthesis workflow for developing CVE to ATT\&CK mapping datasets. It employs retrieval-aware training techniques to enhance the strategic reasoning capabilities of large language models in generating precise policies.
Similarly, \citet{DBLP:journals/corr/abs-2401-00280} comply a dataset of 639 descriptions by extracting tactics, techniques, and sub-techniques from the MITRE ATT\&CK framework and evaluated different models' abilities to interpret process descriptions and map them to corresponding ATT\&CK tactics.

When focusing on log analysis, LLMs also demonstrate strong parsing and analytical capabilities.
For instance, 
\citet{DBLP:journals/corr/abs-2306-01590} assess the capability of ChatGPT in log parsing. They devised appropriate prompts to guide ChatGPT in understanding log parsing tasks and extracting log events/templates from input log messages.
Besides, \citet{DBLP:conf/hpcc/QiHLYFYQSXW23} introduce LogGPT, a log anomaly detection framework based on ChatGPT. Leveraging ChatGPT's natural language understanding capabilities, it explores the potential of transferring knowledge from large-scale corpora to the task of log anomaly detection.
Moreover, \citet{DBLP:journals/corr/abs-2311-14519} explore methodologies for log file analysis using Large Language Models (LLMs) and evaluates the performance of various LLM architectures in the context of application and system log security analysis.

However, the abundance of overlong log files presents a significant hurdle for LLMs, potentially concealing anomalous behavior within truncated logs due to the constrained context length of LLMs.
Moreover, methods that segment and analyze logs fail to provide LLMs with the historical context of logs.
In contrast, we propose the utilization of automated tools to effectively extract user behavior characteristics from extensive log datasets and input them as contextual information into LLMs for assessment.

%!TEX root = ./Audit-LLM.tex
\section{APPROACH}
\begin{figure*}[htbp]
	\centerline{\includegraphics[width=0.95\textwidth]{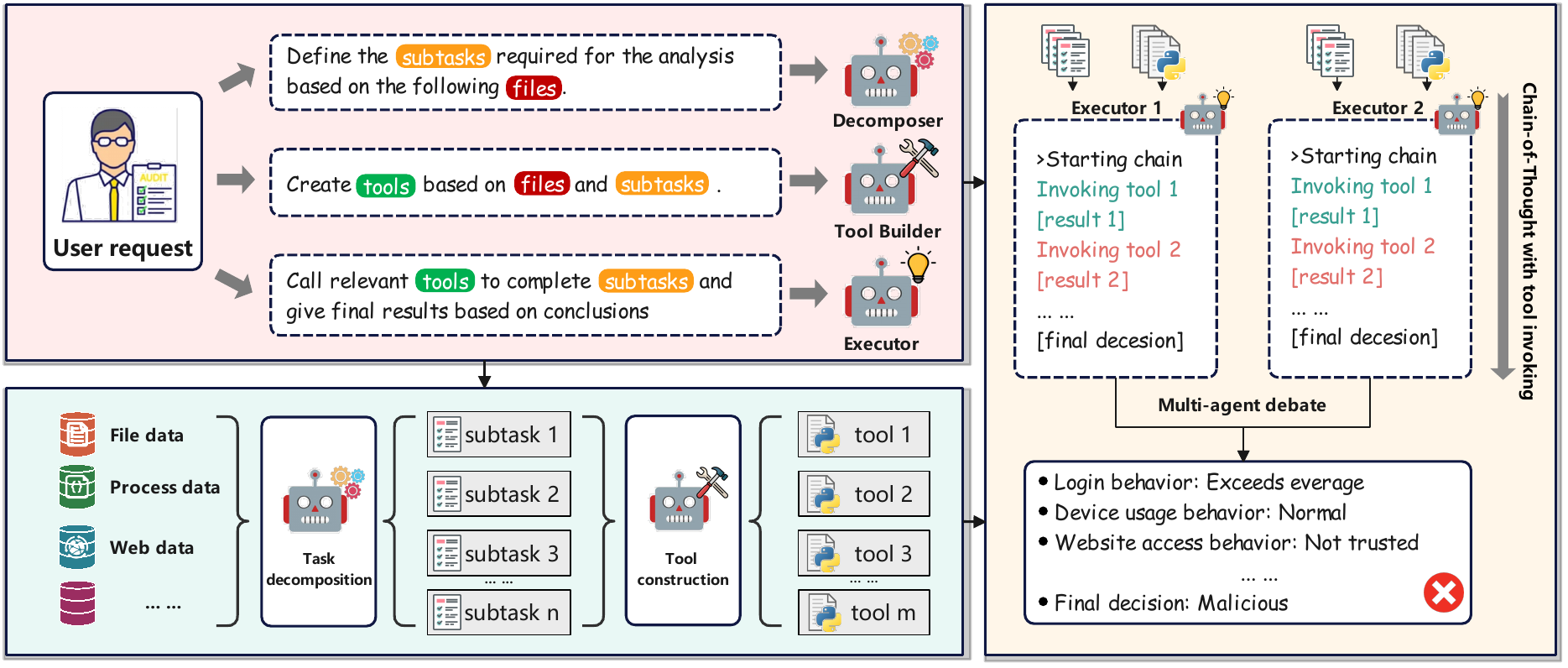}}
	\caption{The framework of Audit-LLM comprises three agents:(i) the Decomposer, tasked with breaking down complex tasks into more manageable sub-tasks via the COT reasoning, (ii) the Tool builder, responsible for creating a suite of task-specific, callable tools;  and (iii) two Executors, dedicated to independently accomplishing the sub-tasks and reach the conclusion consensus by the pair-wise Evidence-based Multi-agent Debate mechanism.}
	\label{frame}
	\label{overall_framework}
\end{figure*}

This section formalizes the task and presents the proposed model, including the framework and module details.

\subsection{Framework}
{We first overview the Audit-LLM framework, as illustrated in Fig.~\ref{overall_framework}.
It is a multi-agent collaboration framework, consisting of three core agents: the Decomposer, the Tool Builder, and the Executor.
Particularly, the Decomposer reformulates the log auditing task into a sequence of more manageable sub-tasks via the CoT reasoning (Sec.~\ref{overview}).
Then, the Tool Builder constructs a set of reusable tools tailored for each sub-task (Sec.~\ref{prompt_construction}).
Ultimately, two independent Executors dynamically invoke tools to accomplish sub-tasks, generating respective results, which are further refined by the pair-wise Evidence-based Multi-agent debate mechanism to culminate in a consensus on the final conclusion (Sec.~\ref{debate}).}

\subsection{Task definition and decomposition}
\label{overview}
\subsubsection{ITD}

% \zheng{
% Formally, consider an information system that accesses a time-ordered log set, $\mathcal{L} = \{a^{u_i}_j \in \mathbb{R} \mid 1 \leq i \leq N, 1 \leq j \leq |u_i|\}$, where each log $a^{u_i}_j$ belongs to an activity type from the set $\mathbb{R}$, encompassing actions like logon events, website visits, file operations, and email contents. 
% %
% The log set is produced by $N$ users, with each user $u_i$ ($i = 1, \ldots, N$) executing a sequence of activities, $S^u=\{a^{u_i}_1, \ldots, a^{u_i}_{|u_i|}\}$, that intertwine with the activities of other users.
% %
% In this light, the goal of ITD is to train a model $\mathbf{M}$ to detect whether this log set $\mathcal{L} $ contains malicious activities $\mathcal{L}_M (\mathcal{L}_M \subseteq \mathcal{L}) $.
% Hence, the detection results $y_{\mathcal{L}}$ can be formulated as:
% \begin{equation}
%    y_{\mathcal{L}}=
%     \begin{cases}
%         \text{benign}, & \text{if } \mathcal{L}_M = \emptyset \\
%         \text{malicious}, & \text{if } \mathcal{L}_M\neq \emptyset
%     \end{cases}, \quad
%     \mathcal{L}_M \leftarrow \mathbf{M}(\mathcal{L})
% \end{equation}
% }

Formally, consider an information system that accesses a time-ordered log set, $\mathcal{L} = \{a^{u_i}_j \in \mathcal{R} \mid 1 \leq i \leq N, 1 \leq j \leq |u_i|\}$, where each log belongs to an activity type from the set $\mathcal{R}$, including actions such as logon events, website visits, file operations, and email contents. 
Here, the log set $\mathcal{L}$ is generated by $N$ users, with each user $u_i$ ($i = 1, \ldots, N$) executing a sequence of activities, $S^{u_i} = \{a^{u_i}_1, \ldots, a^{u_i}_{|u_i|}\}$, interwoven with the activities of others.
For Insider Threat Detection (ITD), the objective is to train a model $\mathcal{M}$ to identify if the log set $\mathcal{L}$ contains malicious activities $\mathcal{L}_M$ ($\mathcal{L}_M \subseteq \mathcal{L}$). 
Suppose all users are independent, ITD essentially can be simplified into analyzing the activity sequence for each user.
Consequently, the detection outcomes $y_{\mathcal{L}}$ are defined as:
\begin{equation}
   y_{\mathcal{L}}=
    \begin{cases}
        \text{benign}, & \text{if } \mathcal{L}_M = \emptyset \\
        \text{malicious}, & \text{if } \mathcal{L}_M \neq \emptyset
    \end{cases}, 
    \mathcal{L}_M \leftarrow \mathcal{M}(\mathcal{L})=\bigcup\limits_{i=1}^{N}\mathcal{M}(S^{u_i})
\end{equation}

% Formally, consider an information system accessible to a set of users $\mathcal{U}=\{u_i|i\in 1,2,\cdots, N\}$, where $N$ represents the total number of users.
% %
% We denote the set of all activities recorded chronologically by the system as $\mathcal{A}=\{a^j|j\in\mathbb{R}\}$, where $\mathbb{R}$ denotes the range of activity types, such as logon events, website visits, file operations, and email contents.
% %
% The activity sequence of user $u\in\mathcal{U}$ is denoted as $S^u=\{a^u_1,a^u_2,\cdots,a^u_{n_u}\}$, where $n_u$ is the length of the activity sequence.
% %
% We define $S^u_m \subseteq S^u$ to represent the set of malicious activities of the user, with $S^u_m=\emptyset$ for benign users.
% %
% The goal of ITD is to train a model $\mathcal{F}$ to predict whether a user $u$ is malicious:
% %
% % \begin{equation}
% %     S^u_m \leftarrow \{a_i^u \in S^u | \mathbbm{1}(\mathcal{F}(a_i^u)=1)\}, \left\{
% %     \begin{aligned}
% %         & benign, S^u_m = \emptyset \\
% %         & malicious, S^u_m \neq \emptyset \\
% %     \end{aligned}
% %     \right.
% % \end{equation}
% \begin{equation}
%    y=
%     \begin{cases}
%         \text{benign}, & \text{if } S^u_m = \emptyset \\
%         \text{malicious}, & \text{if } S^u_m \neq \emptyset
%     \end{cases}, \quad
%     S^u_m \leftarrow \{a_i^u \in S^u \mid \mathcal{F}(a_i^u)=1\} 
% \end{equation}

\subsubsection{CoT for ITD}

The Chain-of-Thought (CoT) is a reasoning mechanism where the Large Language Model (LLM) produces intermediate steps or justifications to reach the final conclusion, thereby improving interpretability and the model's proficiency in tackling intricate tasks. 
For an LLM $\mathcal{M}$ with a COT consisting of $T$ reasoning stages, the iterative refinement of ITD for the log set $\mathcal{L}$ can be expressed as:
\begin{equation}
\begin{cases}
    y_i = \underset{w \in \mathcal{V}}{\text{argmin}}\ \mathcal{M}(w \mid \mathcal{L}, p, y_{i-1}),\\
    y_0 = \emptyset, \quad\quad \text{for} \ i = 1, \ldots, T, 
\end{cases}
\label{cot}
\end{equation}
Here, $\mathcal{V}$ represents the vocabulary of LLM $\mathcal{M}$, with $y_i$ denoting the $i$-th reasoning step's outcome. And the final detection outcomes  $y_{\mathcal{L}}$ is $y_T$. 

% Given a user behavior sequence $S_u \subseteq \mathcal{A}$ to be analyzed, assuming the output result, whether it includes malicious behavior or not, is defined as $y$, the process of an LLM $\mathcal{M}$ with standard prompting can be represented as:
% %
% % \begin{equation}
% %     y=\mathcal{M}(S^u, p).
% % \end{equation}
% \begin{equation}
%     y \leftarrow \text{argmax} \mathcal{M}(S^u, p).
% \end{equation}
% %

However, as previously noted, the log set $\mathcal{L}$ comprises diverse types of activities.
Directly inputting log entries into the LLM may result in overlooking certain types of behaviors.
Thus, we develop an agent named \textbf{Decomposer}, 
tasked with decomposing the tasks of ITD into multiple sub-tasks $\{z^i_{ta}\}_{i=1}^{N_t}$, enabling Audit-LLM to address ITD through a CoT paradigm as:
\begin{equation}
    z^i_{ta}\leftarrow \mathcal{M}(Sample(\mathcal{R}), p_{Deco}), \quad \text{for} \ i = 1, \ldots, N_t,
\end{equation}
where $p_{Deco}$ represents the prompt for constructing the Decomposer, and $Sample()$ denotes the process of sampling three data points from log files as examples for each activity category.
{An examination of a particular activity type may encompass more than one sub-task. 
For instance, the inspection of website access necessitates concurrent checks for the URL, the content of the site, and the potential downloading of malicious payloads. Therefore, the number of sub-tasks is typically greater than or equal to the number of activity types, i.e., $N_t \geq |\mathcal{R}|$.}
% Typically, the number of sub-tasks is greater than or equal to the number of activity types, i.e., $N_t \geq |\mathbb{R}|$. \scy{An examination of a particular activity may encompass more than one sub-task. For instance, the inspection of website access necessitates concurrent checks for the URL, the content of the site, and the potential downloading of malicious payloads.}

To ensure thorough coverage of all potential malicious behavior patterns, we drew inspiration from the field of psychology's concept of ``meta-cognitive dialogues''—a process of self-reflection and iterative improvement~\citep{DBLP:conf/aaaiss/Conway-SmithW24}.
Specifically, we present the Decomposer with slicing of log {activities} and guided its iterative exploration of sub-tasks by continually asking, ``\textit{What additional information do you need to detect threat behaviors?}''
This {strategy} allows the Decomposer to progressively advance until barely surpassing the task's boundaries, i.e., {the requirements of detecting the whole log set}.
% \zheng{
% \begin{equation}
%    min N_t, \quad \text{s.t.} \quad  \bigcup\limits_{i=1}^{N_t}z^i_{ta} \supset  \mathcal{L}  .
% \end{equation}
% }
%

In practice, these sub-tasks guide the Tool Builder agent to construct sub-task-specific tools, which further help the Executor agent to think logically, generating corresponding intermediate results to facilitate the CoT process for ITD (see Eq.\eqref{cot}).

\subsection{Tool development and optimization}
\label{prompt_construction}

% In this section, we demonstrate how the Tool builder constructs invocable tools and enhances their reliability.
%

% The objective of  the Tool Builder agent is to create a set of subtask-specific and reusable tools $\{z_{to}^i\}_{i=1}^{N_t}$ implemented as \textbf{Python functions} to fulfill the completion of sub-tasks $\{z_{ta}^i\}_{i=1}^{N_t}$, 
% % representing a more generic form of CoT that enhances the overall utility and flexibility of the LLMs~\citep{DBLP:journals/corr/abs-2305-17126}.
% which can be formulated as follows:
% %
% \begin{equation}
%     z_{to}^i \leftarrow \mathcal{M}(z_{ta}^i, p_{Tool}), \quad \text{for} \quad i=1,\cdots, N_t,
% \end{equation}
% %
% where $p_{Tool}$ is the prompt of the Tool Builder.
% %
% This process can be further divided into three stages:

The objective of the Tool Builder agent is to construct a collection of sub-task-specific and reusable tools $\{z_{to}^i\}_{i=1}^{N_t}$, implemented as \textbf{Python functions}, to facilitate the completion of sub-tasks $\{z_{ta}^i\}_{i=1}^{N_t}$. The creation process can be formalized as:
\begin{equation}
    z_{to}^i \leftarrow \mathcal{M}(z_{ta}^i, p_{\text{Tool}}), \quad \text{for} \ i = 1, \ldots, N_t,
\end{equation}
where $p_{\text{Tool}}$ denotes the prompt for the Tool Builder. This process is further delineated into three stages:

\subsubsection{Intent recognition}
In this stage, we applied the ``Programming by Example'' (PbE) paradigm~\citep{DBLP:journals/ai/Bauer79}, which streamlines the programming process and reduces complexity by guiding program writing through concrete demonstrations.
Specifically, the demonstrations consist of two components, namely log examples and result examples.
Log examples serve to acquaint the Tool Builder with accessible inputs and enable them to determine the necessary input parameters for the tools.
Result examples aim to align the generated tool with the objectives of the sub-tasks.

\subsubsection{Unit test}
In the next stage, we verify the functionality and accuracy of the generated tools through unit testing, ensuring that the agent can seamlessly invoke these tools and obtain accurate results.
We begin by manually entering the necessary parameters for the tools to validate their functionality.
Subsequently, each tool undergoes three invocation tests by the LLM for every sub-task. 
Upon successful extraction of parameters from the logs and their placement in the intended positions by the LLM, the tool is incorporated into the toolkit for subsequent utilization.
While any tool triggers errors during this process, we document the error details and utilize this information to guide the LLM in reconstructing the faulty tool.

\subsubsection{Tool decoration}
Although unit testing has demonstrated the reliability of an agent's invocation of individual tools, the presence of multiple tools can still potentially lead to error invocation.
Thus, the final stage involves enhancing the validated tools with decoration.
The decoration process includes two main aspects: code documentation and output restructuring.
Code documentation provides contextual information at the beginning of each tool, accurately describing its functionality to prevent missteps during agent calls. 
Additionally, we refine the output formats to ensure that the agent can better understand the meaning of the results obtained from using the tools in natural language terms.

In practice, the Tool Builder completes the process of tool development and optimization, and only needs to be performed once for each sub-task.
The resulting tools {$\{z_{to}^i\}_{i=1}^{N_t}$} can be reused for all instances of ITD, thereby reducing the usage costs of Audit-LLM.
%
% Furthermore, Python-based tools are a more generic form of CoT, enhancing the overall utility and flexibility of the LLMs~\citep{DBLP:journals/corr/abs-2305-17126}.
%

\begin{figure}
	\includegraphics[width=0.45\textwidth]{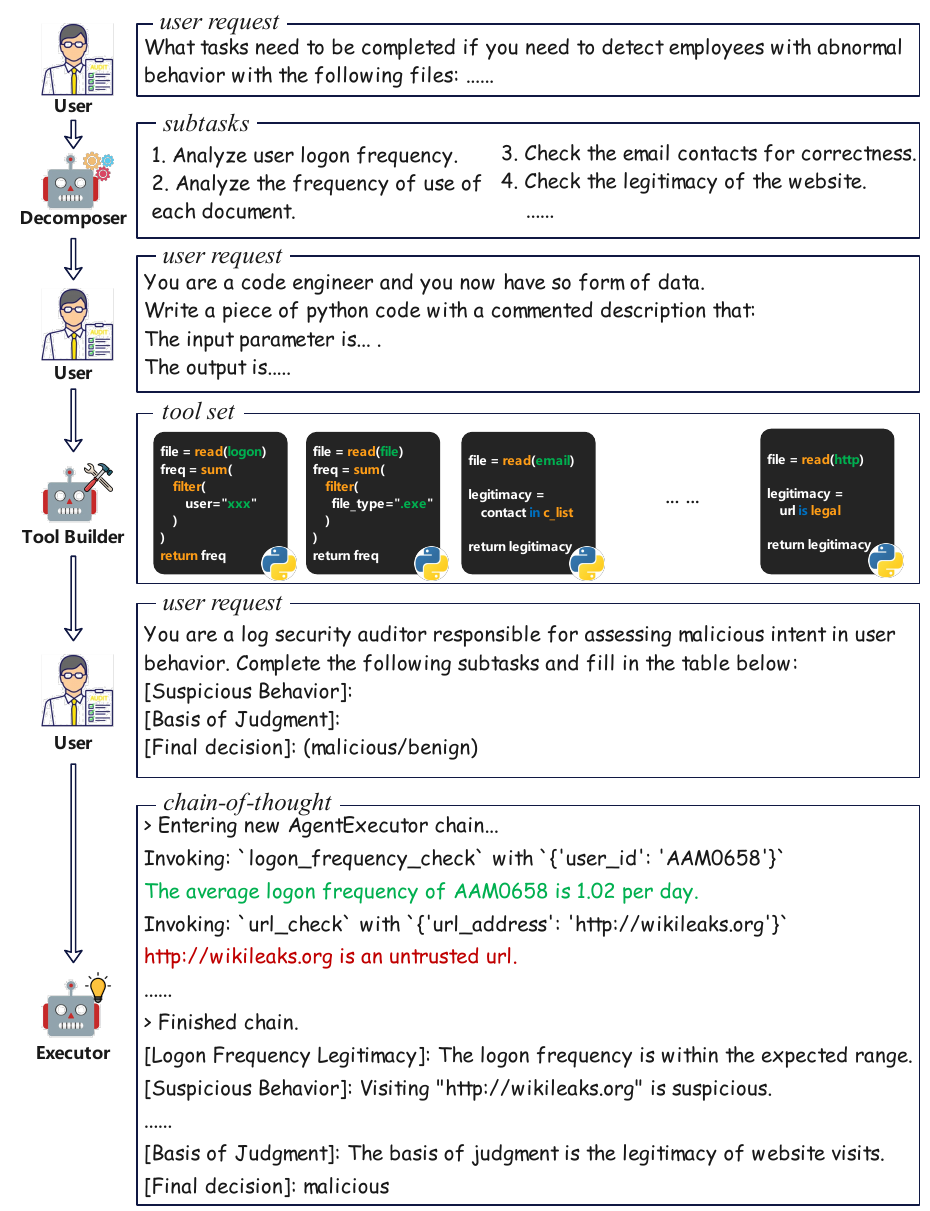}
	\caption{An example of building different agents via prompts. The specific details are omitted for the sake of brevity.}
	\label{prompt}
\end{figure}

\begin{algorithm}[t]
\small
	\caption{The procedure of pair-wise Evidence-based Multi-agent debate.}
	\label{Algorithm1}
	\KwIn{
		Given a task set comprising $n$ sub-tasks $\{z_{ta}^i\}_{i=1}^{N_t}$ and a set of tools $\{z_{to}^i\}_{i=1}^{N_t}$.
	}
	\KwOut{Conclusion regarding the malignancy of the log set $y_\mathcal{L}^*$, along with the rationale behind the judgment.}
	\BlankLine
	Initialize two Executors $E_a$ and $E_b$ with prompt $P_{Exec}$.
	$E_a, E_b = \mathcal{M}(P_{Exec})$
	\hfill 
	\\
	$\{results\}_{E_a}^0 \leftarrow E_a(\{z_{ta}^i\}_{i=1}^{N_t}, \{z_{to}^i\}_{i=1}^{N_t})$
	\\
	$\{results\}_{E_b}^0 \leftarrow  E_b(\{z_{ta}^i\}_{i=1}^{N_t}, \{z_{to}^i\}_{i=1}^{N_t})$
	\\
	Set the maximum number of iterates of debate $N_{debate}$
	\\
	\For {$i=1$ to $N_{debate}$}{
		\If{$\{y_\mathcal{L}\}_{E_a}^i=\{y_\mathcal{L}\}_{E_b}^i$}{
				break
			}
		\Else{
			$\{results\}_{E_a}^i \leftarrow E_a(\{results\}_{E_a}^{i-1},\{results\}_{E_b}^{i-1})$
			\\
			$\{results\}_{E_b}^i \leftarrow E_b(\{results\}_{E_a}^i,\{results\}_{E_b}^{i-1})$
			}
		\textbf{end if}
	}	
	\textbf{end for} 
	\\
	$y_\mathcal{L}^* \leftarrow E_a(\{results\}_{E_a}^{N_{debate}}, \{results\}_{E_b}^{N_{debate}})$
	\\
	return: $y_\mathcal{L}^*$
\end{algorithm}

\subsection{Task execution}
\label{debate}

The principal role of the Executor is to complete the sub-tasks generated by the Decomposer by invoking tools in a CoT manner.
Hence, the intermediate results can be formulated as:
\begin{equation}
    z_{ex}^i \leftarrow \mathcal{M}(z_{ta}^i, Invoke(z_{to}^i), p_{Exec}), \ \text{for} \ i = 1, \ldots, N_t,
\end{equation}
where the invoke function $Invoke(\cdot)$  returns the outputs invoked by tools.
After that, the Executor synthesizes the tool-derived results to form the ultimate conclusion $y_\mathcal{L}$ i.e., $y_\mathcal{L}=\bigcap\limits_{i=1}^{N_t} z_{ex}^i $.
We illustrate the collaborative process of all agents in Fig.~\ref{prompt}.
To manage output diversity, we instruct the Executor to categorize responses into predefined formats, e.g., sub-task results, anomalous behaviors, and final conclusions.
%
% If the output conforms to the pre-defined format, the response is directly parsed; otherwise, the Executor is prompted to reformat the response to meet the specific requirements.
% \vspace{-1pt}

Although generally effective, the Executor occasionally generates final conclusions that may contradict the results of sub-tasks.
For instance, consider the scenario where the Executor correctly invokes the \textit{website legitimacy verification} tool and identifies some malicious activities, e.g., accessing suspicious sites or downloading malicious payloads.
Nevertheless, the final conclusion synthesized by LLM may still categorize this log set as benign, resulting in what is known as faithfulness hallucination in LLM~\citep{DBLP:journals/corr/abs-2311-05232}.
This phenomenon can be attributed to the intrinsic event error from LLM~\citep{DBLP:journals/corr/abs-2402-07401}, wherein the generated explanations misrepresent source events.
%
% In this context, among the nine intrinsic and extrinsic errors defined by \citet{DBLP:journals/corr/abs-2402-07401}, the primary challenge encountered by Audit-LLM predominantly lies in intrinsic event error, i.e., the generated explanation misrepresents events mentioned in the source.
%

% The iterative refinement paradigm, exemplified by self-consistency~\citep{DBLP:conf/nips/MadaanTGHGW0DPY23,DBLP:conf/iclr/0002WSLCNCZ23}, incurs significant computational overhead, as each Executor is required to traverse an entire CoT process, involving the invocation of multiple tools to complete sub-tasks. 
% %
% \scy{Additionally, existing multi-agent debate methods either follow a majority-rule approach or introduce an additional agent as a ``judge.'' 
% %
% Either way, at least three agents are required, resulting in a significant increase in token usage.}
%
To address this issue, we propose a pair-wise Evidence-based Multi-Agent Debate (EMAD) mechanism, mimicking the human debate process to help LLM review each reasoning step.
Particularly, two Executors are designated as the proponent and opponent, with their respective reasoning processes forming the basis of their claims. 
During each debate round, these claims are presented to the opposing Executor for scrutiny, which helps each Executor to update their respective conclusion.
Through multiple rounds of debate, a comprehensive and accurate conclusion consensus can be achieved.

%potential thought processes or {information source} for further examination by LLM.
%During debate, the individual response of each Executor, i.e., the thought process and conclusions, are combined and presented as context to the other Executor.}
%
% Following the completion of the CoT process by these two Executors, a round of debate between them is initiated.
%
%Subsequently, two independent Executors will review each other’s thought processes, including the correctness of tool usage and the results of sub-tasks, as well as whether the final conclusion aligns with the results of the sub-tasks.
%We iteratively repeat this debate procedure over multiple rounds for the conclusion coverage.

Formally, the process of EMAD is outlined in Algorithm~\ref{Algorithm1}.
%EMAD employs two Executors to identify errors and engage them in a debate until a consensus is reached.
%
{Given a sub-task set $\{z_{ta}^i\}_{i=1}^{N_t}$ as well as its corresponding tool set $\{z_{to}^i\}_{i=1}^{N_t}$} 
%generated by the Decomposer and the Tool Builder
, the Executors $E_a$ and $E_b$ are 
%tasked with producing conclusions regarding the malignancy of the log entry, accompanied by the reasoning behind the judgment.
%We 
first initialized with prompts in line 1.
Subsequently, in lines 2-3, $E_a$ and $E_b$ generate initial results $\{results\}_{E_a}^0$ and $\{results\}_{E_b}^0$, respectively.
%by invoking relevant tools and completing sub-tasks.
%
Here, $\{results\}$ are constituted by the collection of the final conclusion and its intermediate results of the Executor, i.e., $\{y_\mathcal{L}, z_{ex}^1, \cdots, z_{ex}^{N_t}\}$.
For each $i$-th iteration of debate, $E_a$ and $E_b$ review the opposing reasoning process in the previous $(i-1)$-th iteration (i.e., $\{results\}_{E_a}^{i-1}$ for $E_a$ and $\{results\}_{E_b}^{i-1}$ for $E_b$) to update  their respective conclusions in lines 5-12.
To ensure the most accurate feedback, the two Executors continue their discussion until they reach a mutual agreement on the result set. 
{Finally, merging the final results from both Executors formulates the ultimate conclusion, as indicated in line 13.}
To prevent endless debates, we also set a fixed number of iterations in line 4.

%!TEX root = ./Audit-LLM.tex
\section{Experimental Setup}

Within this section, we present a comprehensive overview of the experimental configuration, encompassing details about the experimental environment, datasets, and baselines.
First, we present detailed information about the experimental configuration and discuss the research questions in Sec.~\ref{configuration}.
Second, we provide an in-depth description of the datasets used in our study in Sec.~\ref{dataset}.
Finally, we summarize the baseline models in Sec.~\ref{baseilnes}

\subsection{Research questions and experimental configuration}
\label{configuration}

\subsubsection{Research questions}

We list several research questions to guide the experiments and verify the effectiveness of our proposal.
\begin{enumerate}[nosep, align=left, leftmargin=*]
	\item[\bf{RQ1}:] Can the proposed Audit-LLM achieve better performance than state-of-the-art baselines for log-based ITD?
	
	\item[\bf{RQ2}:] Which component contributes more to improving model performance?
	
	\item[\bf{RQ3}:] When employing diverse LLMs as the base models, how does the performance of Audit-LLM vary?
	
	\item[\bf{RQ4}:] What do the responses generated by Audit-LLM look like? How interpretable are they? 
	
	\item[\textbf{RQ5}:] How does Audit-LLM perform in real-world system environments?
	
	\item[\bf{RQ6}:] What are the time and economic implications of using online LLM APIs like ChatGPT and ZhipuAI?
\end{enumerate}

\subsubsection{Experimental configuration}

The experiments are conducted with an Intel Xeon(R) Gold 5218R CPU, 256 GB of RAM, and four Nvidia RTX A6000 (48 GB) GPUs.
The agent in Audit-LLM is developed based on LangChain~\citep{langchain}, and we conduct our method in the Python 3.10.14 environment.
We build our Audit-LLM framework based on multiple large language models, including a snapshot of gpt-3.5-turbo-0125 released by Openai~\citep{openai}, 
To facilitate the reproduction of the results in this paper, we share the code and data used to obtain these results on \url{https://anonymous.address/}.
Additionally, we have shared the prompt we developed in the LangChain prompt hub under the identifier ``anonymous-id''.

\subsection{Datasets}
\label{dataset}

\begin{table}[t]
	\begin{spacing}{1.2}
	\renewcommand{\arraystretch}{1.1}
	\centering
	\setlength{\tabcolsep}{2.6mm}{
	\small
	\caption{Statistics of the datasets used in our experiments.}
	\label{statistic}
	\begin{tabularx}{0.47\textwidth}{lllX}
		
		\hline
	Dataset                 & \multicolumn{1}{l}{CERT r4.2}                           & \multicolumn{1}{l}{CERT r5.2}                          & \multicolumn{1}{l}{PicoDomain}                                                                                     \\ \hline
	Duration                & \multicolumn{2}{c}{18 months}                                                                                    & 3 days                                                                                                             \\
	\# employees            & 930                                                     & 1,901                                                  & 15                                                                                                                 \\
	\# insiders             & 70                                                      & 99                                                     & 1                                                                                                                  \\
	\# benign entries    & 32,762,906                                              & 79,846,358                                             & 537,840                                                                                                            \\
	\# malicious entries & 7,316                                                   & 10,306                                                 & 80                                                                                                                 \\
	Data sources            & \multicolumn{2}{l}{\begin{tabular}[c]{@{}l@{}}logon, email, \\ device, http, file, \\ psychometric\end{tabular}} & \multicolumn{1}{l}{\begin{tabular}[c]{@{}l@{}}file, system, \\ network, auth, \\ anomalies\end{tabular}} \\
		\hline
	\end{tabularx} }
\end{spacing}
\end{table}%

To conduct robust and convincing experiments, we utilize three publicly accessible insider threat datasets, namely, CERT r4.2, CERT 5.2~\citep{CERT}, and PicoDomain~\citep{DBLP:journals/corr/abs-2008-09192}.
The data statistics are summarized in Table~\ref{statistic}.

The CERT datasets, provided by Carnegie Mellon University~\citep{DBLP:conf/sp/GlasserL13} in this work, are widely recognized in log-based insider threat detection~\citep{DBLP:journals/compsec/XiaoZZLDL24,DBLP:conf/ipccc/SunY22,DBLP:conf/ccs/LiuWZJXM19,DBLP:journals/corr/abs-2403-09209,GONCALVES2024103944}.
On one hand, CERT r4.2 includes activity logs from 1,000 users and 1,003 computers, while CERT r5.2 simulates an organization with 2,000 employees over 18 months.
Both datasets encompass diverse multi-source activity logs, including user \textit{login/logoff events}, \textit{emails}, \textit{file access}, \textit{website visits}, \textit{device usage}, and \textit{organizational structure} data.
Each malicious insider in the CERT dataset is categorized into one of four prevalent insider threat scenarios: data exfiltration, intellectual property theft, and IT sabotage.

On the other hand, PicoDomain consists of detailed Zeek logs spanning 3 days, collected from a simulated small-scale network where APT attacks occurred during the last two days.
The data sources within PicoDomain can be broadly classified into 5 groups: file logs (\textit{files}, \textit{smb\_files}), system logs (\textit{dhcp}, \textit{hosts}, \textit{services}), authentication logs (\textit{kerberos}, \textit{ntlm}), and anomaly detection logs (\textit{weird}).

\subsection{Baseline methods}
\label{baseilnes}

\subsubsection{Baseline methods}

For all LLM-based models discussed, we employ the same base LLM, i.e., the snapshot of gpt-3.5-turbo-0125, to fairly compare their performance.
Due to severe class imbalance in the CERT dataset, which impedes DL-based methods from effectively capturing features of minority classes, we implemented an under-sampling approach following~\citet{DBLP:journals/compsec/XiaoZZLDL24}, restricting the number of benign class samples to below 20,000. 
%
%This under-sampling strategy is specifically tailored for DL-based methods, acknowledging their challenge in adequately capturing minority class features from highly imbalanced data.
%
Here, we list a series of state-of-the-art baselines for comparisons with our proposal in this paper:
\begin{enumerate}[nosep, align=left, leftmargin=*]
	\item[\bf{LogGPT} \citep{DBLP:conf/hpcc/QiHLYFYQSXW23}:] LogGPT utilizes LLMs for log auditing, extracting structured data from raw logs audited by ChatGPT.
	
	\item[\bf{LogPrompt} \citep{DBLP:conf/icse/0001TMYZY24}:] LogPrompt enhances zero-shot log auditing using LLMs and advanced prompting techniques, employing their top-performing CoT prompt.
	
	\item[\bf{LAN} \citep{DBLP:journals/corr/abs-2403-09209}:] LAN employs graph structure learning to adaptively construct user activity graphs, addressing data imbalance with a hybrid predictive loss.
	
	\item[\bf{DeepLog} \citep{DBLP:conf/ccs/Du0ZS17}:] DeepLog treats log entries as sequential natural language, utilizing Long Short-Term Memory to detect anomalies.
	
	\item[\bf{LMTracker} \citep{DBLP:journals/ijon/FangWFH22}:] LMTracker uses event logs to construct heterogeneous graphs and apply unsupervised algorithms to detect malicious behavior.
	
	\item[\bf{CATE} \citep{DBLP:journals/compsec/XiaoZZLDL24}:] CATE uses convolutional attention and a transformer encoder for log statistical and sequential analysis.
\end{enumerate}
In our comprehensive model performance evaluation, we covered baselines using LLMs, LSTM for sequential data, GNNs for graph structures, and pre-trained Transformers for insider threat detection.

%!TEX root = ./Audit-LLM.tex

\section{Results and analysis}

\subsection{Overall performance}
\label{overall}

\begin{table*}[t]
	\captionsetup{justification=justified}
	\centering
%	\large
	\caption{Performance comparison of Audit-LLM with six baselines for insider threat detection. The best and second-best results are boldfaced and underlined, respectively. An upward arrow ($\uparrow$) indicates the higher the better, and a downward arrow ($\downarrow$) indicates the lower the better.}
	\begin{spacing}{1.3}
	\begin{tabular}{lcccccccccccc}
		\hline
		\multicolumn{1}{c}{\multirow{2}{*}{\textbf{Model}}} & \multicolumn{4}{c}{\textbf{CERT r4.2}}                                     & \multicolumn{4}{c}{\textbf{CERT r5.2}}                                     & \multicolumn{4}{c}{\textbf{PicoDomain}}                                    \\ \cmidrule(r){2-5} \cmidrule(r){6-9} \cmidrule(r){10-13}
		\multicolumn{1}{c}{}                       & \textbf{Prec $\uparrow$}          & \textbf{DR $\uparrow$}             & \textbf{FPR $\downarrow$}            & \textbf{Acc $\uparrow$}            & \textbf{Prec $\uparrow$}           & \textbf{DR $\uparrow$}              & \textbf{FPR $\downarrow$}           & \textbf{Acc $\uparrow$}             & \textbf{Prec $\uparrow$}           & \textbf{DR $\uparrow$}              & \textbf{FPR $\downarrow$}            & \textbf{Acc$\uparrow$}            \\ \hline
		DeepLog                                             & 0.684                    & 0.715                  & 0.335                     & 0.743                   & 0.728                    & 0.776                  & 0.264                     & 0.801                   & 0.728                    & 0.746                  & 0.752                     & 0.754                  \\
		LMTracker                                           & 0.782                    & 0.829                  & 0.217                     & 0.856                   & 0.765                    & 0.794                  & 0.216                     & 0.821                   & \underline{0.902}              & \underline{0.918}            & \underline{0.095}               & \underline{0.897}            \\
		CATE                                                & 0.885                    & 0.892                  & 0.289                     & 0.928                   & 0.893                    & 0.906                  & 0.324                     & 0.932                   & -                        & -                      & -                         & -                      \\
		LAN                                                 & 0.876                    & 0.886                  & \underline{0.142}               & \underline{0.934}             & 0.883                    & 0.891                  & \underline{0.099}               & 0.902                   & 0.874                    & 0.886                  & 0.122                     & 0.862                  \\
		LogPrompt                                           & 0.861                    & 0.875                  & 0.324                     & 0.841                   & 0.852                    & 0.862                  & 0.328                     & 0.873                   & 0.771                    & 0.804                  & 0.293                     & 0.795                  \\
		LogGPT                                              & \underline{0.911}              & \underline{0.914}            & 0.184                     & 0.926                   & \underline{0.905}              & \underline{0.907}            & 0.116                     & \underline{0.918}             & 0.802                    & 0.824                  & 0.196                     & 0.834                  \\ \hline \hline
		Audit-LLM                                           & \textbf{0.943}           & \textbf{0.958}         & \textbf{0.037}            & \textbf{0.961}          & \textbf{0.941}           & \textbf{0.956}         & \textbf{0.039}            & \textbf{0.959}          & \textbf{0.914}           & \textbf{0.927}         & \textbf{0.067}            & \textbf{0.931}         \\ \hline 
	\end{tabular}
	\end{spacing}
	\label{OP}
\end{table*}

To answer \textbf{RQ1}, we evaluate the performance of our proposed Audit-LLM and six competitive baselines for the insider threat detection task on three public datasets.
We present the results of the involved models in Table~\ref{OP}. 
Among these metrics, the higher the precision, detection rate, and accuracy, the better overall performance.
Contrarily, the lower the False Positive Rate (FPR), the fewer false positives that cause false alarms.

Generally, comparing the model performance on CERT r4.2 against that on CERT r5.2, we can observe that the model mostly performs relatively better on the former than on the latter dataset.
It could be explained by the fact that the r4.2 version of CERT is a ``dense needle'' dataset that contains more insiders and malicious activities than the r5.2 version.
The more severe category imbalance problem results in difficulties for the model to classify the activities correctly.
In particular, our proposed Audit-LLM performs the best among the models, with a noticeable performance improvement over the other six baselines.
For instance, on the CERT r4.2 dataset, Audit-LLM presents an improvement of 21.8\%, 10.5\%, 3.3\%, 2.7\%, 12\%, and 3.5\% in terms of accuracy against the DeepLog, LMTracker, CATE, LAN, LogPrompt, and LogGPT models, respectively.
% , while shows a corresponding 15.8\%, 13.8\%, 2.7\%, 6.7\%, 8.6\%, and 4.1\% on the CERT r5.2 dataset.
%
These overwhelming results indicate that our Audit-LLM leads to consistent gains across different datasets.

For all LLM-based methods, namely LogPrompt, LogGPT, and Audit-LLM, their performance on the CERT dataset is significantly better than on the PicoDomain dataset.
%
% For example, LogPrompt achieves accuracy rates of 0.841 and 87.3\% on CERT r4.2 and r5.2, respectively, but drops to 79.5\% on the PicoDomain dataset. 
%
A similar trend is observed with Audit-LLM, where the accuracy decreases by up to 3\% compared to its performance on the CERT dataset.
The reason is that, compared to CERT which includes more user behavior logs such as email communications and accessed website content, PicoDomain comprises more traffic and system activity logs. 
LLM inherently possesses strong natural language understanding capabilities, thus enabling it to effectively leverage email content or website content summaries for insider threat detection on the CERT dataset.
Note that CATE~\citep{DBLP:journals/compsec/XiaoZZLDL24} integrates organizational structure information and psychological data (user profile data) for each user into a unified table, which is an integral part of the model and therefore not reproducible in terms of performance within PicoDomain.

When zooming in on the False Positive Rate (FPR), it can be observed that the FPRs of Audit-LLM are 3.7\%, 3.9\%, and 6.7\% on CERT r4.2, r5.2, and Picodomain, respectively, lower than all the baselines by at least 1.47\%, 0.8\%, and 2.8\% on three datasets.
These results indicate that Audit-LLM can well reduce the number of false positives, which has great value when the investigation budget is finite.
In addition, LMTracker demonstrates the best performance in the baseline on PicoDomain, achieving 90.2\% precision, 91.8\% detection rate, 9.5\% false positive rate, and 89.7\% accuracy. 
This could be attributed to LMTracker's use of heterogeneous graphs to model relationships between computers and users, specifically focusing on designing models for lateral movement.

\subsection{Ablation study in Audit-LLM}
\label{ablation}

\begin{table*}[t]
	\captionsetup{justification=justified}
	\setlength\tabcolsep{3.5pt}
	\centering
%	\small
	\caption{Ablation study results of Audit-LLM on CERT r4.2, r5.2, and Picodomain. The biggest drop in each column is appended with $\downharpoonright$.}
	\begin{spacing}{1.35}
		\begin{tabular}{lcccccccccccc}
			\hline
			\multicolumn{1}{c}{\multirow{2}{*}{\textbf{\begin{tabular}[c]{@{}c@{}}Model\\ Variants\end{tabular}}}} & \multicolumn{4}{c}{\textbf{CERT r4.2}}                                     & \multicolumn{4}{c}{\textbf{CERT r5.2}}                                     & \multicolumn{4}{c}{\textbf{PicoDomain}}                                    \\ \cmidrule(r){2-5} \cmidrule(r){6-9} \cmidrule(r){10-13}
			\multicolumn{1}{c}{}                                        & \textbf{Prec}  & \textbf{DR}    & \textbf{FPR}   & \textbf{Acc}   & \textbf{Prec}  & \textbf{DR}    & \textbf{FPR}   & \textbf{Acc}   & \textbf{Prec}  & \textbf{DR}    & \textbf{FPR}   & \textbf{Acc}   \\ \hline
			Vanilla (GPT-3.5-turbo)                                     & 0.659          & 0.673          & 0.342          & 0.662          & 0.629          & 0.631          & 0.264          & 0.621          & 0.608          & 0.616          & 0.402          & 0.610          \\ \hline
			Audit-LLM w/o CoT                                           & 0.693$\downharpoonright$          & 0.712$\downharpoonright$          & 0.292$\downharpoonright$          & 0.706$\downharpoonright$          & 0.675$\downharpoonright$          & 0.682$\downharpoonright$          & 0.316$\downharpoonright$          & 0.678$\downharpoonright$          & 0.622$\downharpoonright$          & 0.634$\downharpoonright$          & 0.381$\downharpoonright$          & 0.627$\downharpoonright$          \\
			Audit-LLM w/o Decomp                                        & 0.846          & 0.856          & 0.172          & 0.834          & 0.823          & 0.821          & 0.189          & 0.812          & 0.824          & 0.834          & 0.192          & 0.821          \\
			Audit-LLM w/o tools                                         & 0.876          & 0.883          & 0.137          & 0.871          & 0.867          & 0.877          & 0.128          & 0.863          & 0.841          & 0.852          & 0.313          & 0.835          \\
			Audit-LLM w/o EMAD                                          & 0.921          & 0.935          & 0.069          & 0.946          & 0.932          & 0.941          & 0.054          & 0.948          & 0.904          & 0.924          & 0.072          & 0.923          \\ \hline \hline
			Audit-LLM (original)                                        & \textbf{0.943} & \textbf{0.958} & \textbf{0.037} & \textbf{0.961} & \textbf{0.941} & \textbf{0.956} & \textbf{0.039} & \textbf{0.959} & \textbf{0.914} & \textbf{0.927} & \textbf{0.067} & \textbf{0.931} \\ \hline
		\end{tabular}
	\end{spacing}
	\label{AS}
\end{table*}

For \textbf{RQ2}, we perform an ablation study by comparing Audit-LLM with its variants to analyze the effectiveness of each component.
Specifically, we produce five variants for comparison: 
(1) ``Vanilla'' that removes all agent involvement, providing only the most basic task prompts to the LLM, 
(2) ``Audit-LLM w/o CoT'' that allows the LLM to make one-step decisions without employing reasoning through CoT, 
(3) ``Audit-LLM w/o Decomp'' that removes the Decomposer, thereby not delineating the subtasks that need to be accomplished, and relies solely on tools to assist the Executor in making decisions, 
(4) ``Audit-LLM w/o tools'' that removes the use of tools and allows the Executor to make decisions based on the log entries within the current input,
(5) ``Audit-LLM w/o EMAD'' that removes the evidence-based multi-agent debate process, adopting the initial decision made by the Executor.
The results are presented in Table~\ref{AS}.

From Table~\ref{AS}, we can observe that removing any component in Audit-LLM leads to a decrease in the performance, indicating that all components in Audit-LLM contribute to the model performance.
Besides ``Vanilla'', the largest impact on model performance comes from removing the CoT process, indicating that conducting log auditing in a single step may overlook malicious behaviors, whereas decomposing it into sub-tasks allows for a more comprehensive log auditing.
Moreover, the ``Audit-LLM w/o EMAD'' shows minimal impact on model performance degradation.
This may be attributed to the assistance of CoT, which mitigates the complexity of the ITD task, consequently diminishing the potential for LLM hallucinations.
Thus, EMAD corrects only a small amount of unfaithful hallucinations.

By comparing ``Audit-LLM w/o Decomp'' with ``Audit-LLM (original)'', we observe a notable decline in model performance.
Specifically, ``Audit-LLM w/o Decomp'' shows a decrease in accuracy by 12.7\%, 14.7\%, and 11\% on CERT r4.2, r5.2, and PicoDomain, respectively.
These results suggest that the Decomposer can break down ITD into more manageable sub-tasks based on multiple log data sources. 
By generating a series of intermediate steps before reaching the final result, this approach enhances the LLM's reasoning capabilities.
In addition, ``Audit-LLM w/o tools'' loses the performance competition to the original Audt-LLM on three datasets.
For example, "Audit-LLM w/o tools" exhibits increased false positive rates compared to "Audit-LLM (original)" on CERT r4.2, r5.2, and PicoDomain, with increases of 28.7\%, 28.9\%, and 22.6\%, respectively.
This phenomenon demonstrates that the contextual information provided by tools helps the Executor to eliminate some anomalies within the input window that appear suspicious but are actually benign behaviors.

\subsection{Impact of base LLMs}
\label{base}

\begin{figure}[t]
	\centering
	\begin{subfigure}[t]{\columnwidth}
		\includegraphics[width=\textwidth]{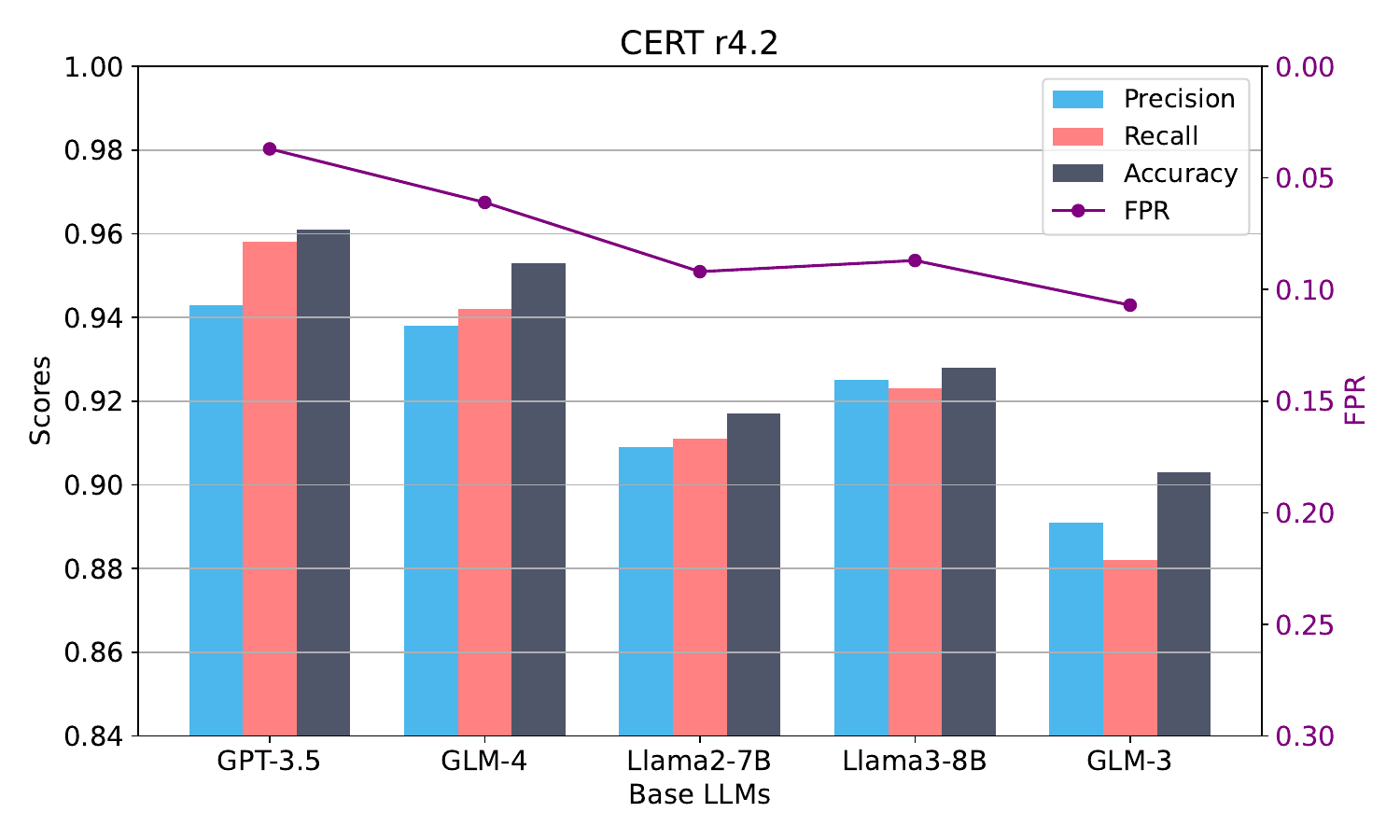}
		\vspace*{-1.5\baselineskip}
		\caption{Performance on CERT r4.2s.}
		\label{emb:distance1}
	\end{subfigure}
	\begin{subfigure}[t]{\columnwidth}
		\includegraphics[width=\textwidth]{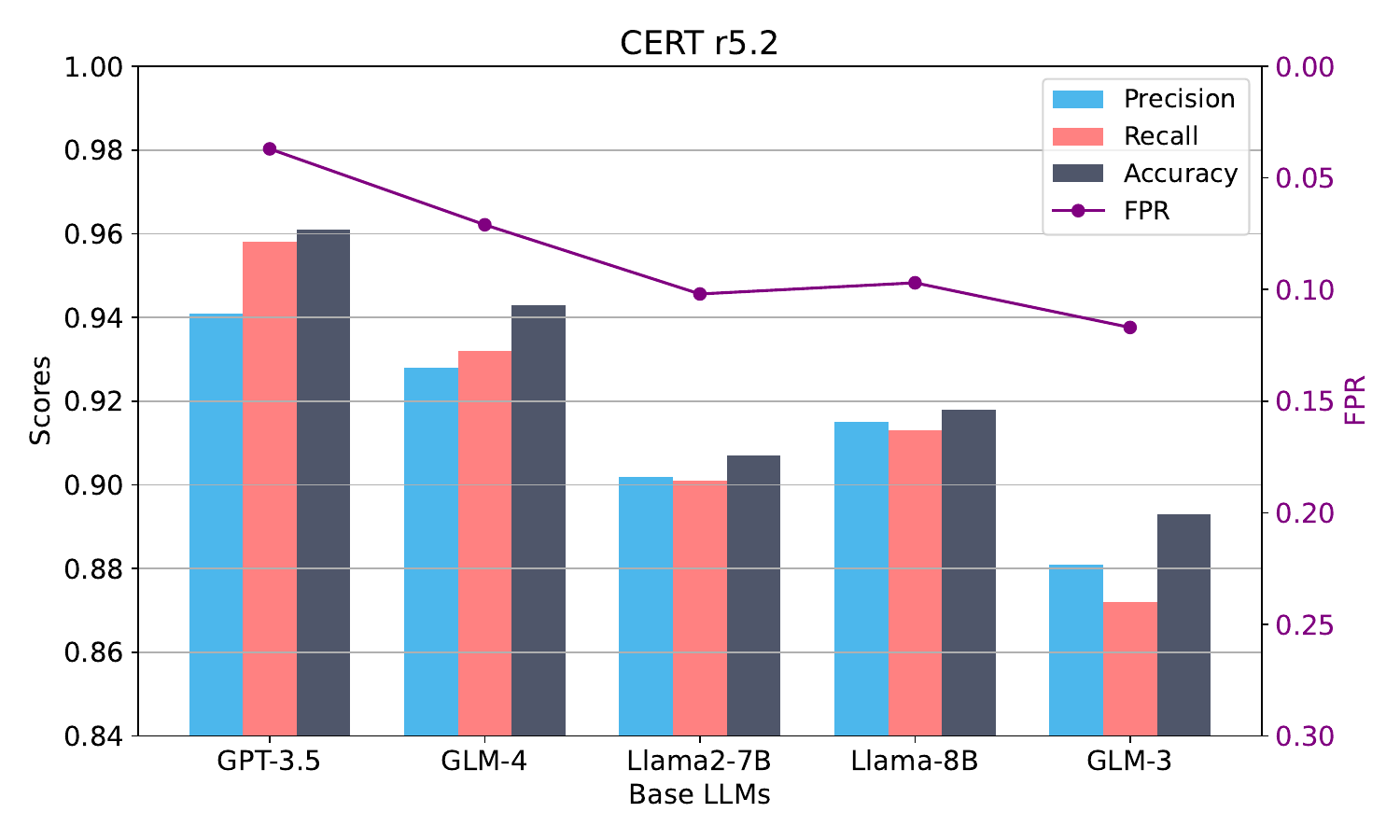}
		\vspace*{-1.5\baselineskip}
		\caption{Performance on CERT r5.2.}
		\label{emb:distance2}
	\end{subfigure}  
	%	\begin{subfigure}[t]{0.99\columnwidth}
		%		\includegraphics[width=\textwidth]{figures/embedding_space}
		%		%\vspace*{-.5\baselineskip}
		%		\caption{Antonym-aware embedding space.}
		%		\label{emb:space}
		%	\end{subfigure}  
	\caption{Performance of Audit-LLM with different base LLMs.}
	\label{emb:EMB}
	%	\vspace*{-2mm}
\end{figure}

Due to Audit-LLM's significant dependence on the natural language understanding and reasoning abilities of the base LLM, it is of interest to investigate the impact of various base LLMs on model performance.
To answer \textbf{RQ3}, we employed GPT-3.5-turbo, Llama2-7B, Llama3-8B-instruct, ChatGLM3-6B, and ChatGLM4 (Zhipuai)~\citep{zhipuai} for evaluation.
For simplicity, we focus our examination solely on the performance of these LLMs on CERT r4.2 and r5.2.
The results are shown in Fig.~\ref{base}.

Overall, we find that the performance of Audit-LLM maintains robust performance across various base LLMs, indicating that the proposed Audit-LLM does not impose stringent requirements on the language comprehension skills of the base LLMs.
The best performance was obtained with GPT-3.5-turbo-0125 on CERT r4.2.
Particularly, GPT-3.5-turbo-0125 shows a maximum accuracy gap of 5.8\% compared to other base language models on the CERT r4.2 dataset, with a minimum gap of 0.08\%.
This phenomenon may be attributed to several factors:
Firstly, GPT-3.5-Turbo-0125 utilized through API calls incorporates significantly higher parameter counts compared to locally executable base language models such as LLama2-7B and ChatGLM3-6B. 
This implies that GPT-3.5-Turbo possesses enhanced learning and inference capabilities when handling large-scale datasets, as it can leverage more parameters to capture complex patterns and features within the dataset.
Secondly, as mentioned, our CoT process is implemented using the LangChain Python library, which was originally tailored for adaptations to the GPT series. 
Such optimizations encompass model invocation methods, data processing, and task-specific adaptations, thereby enhancing the precision of GPT-3.5-Turbo-based agents during the execution of the Chain of Thought (CoT) process and the invocation of tools.

\subsection{Analysis of generated responses}
\label{respnose}

\begin{table}[ht]
%	\small
	\centering
	\caption{The responses generated by Audit-LLM across three scenarios.}
	\label{response}
	\begin{tabularx}{\columnwidth}{X}
		\toprule
		\multicolumn{1}{c}{\textbf{scenario: Benign}} \\ \hline
		\textbf{Logon}: Logon frequency is within normal range.\\
		\textbf{Website}: Websites visited are trusted.\\
		\textbf{Drive}: Driver usage within normal range.\\
		\textbf{Suspicious}: No suspicious behavior detected.\\
		\textbf{Basis of Judgment}: Logon, website, and driver.\\
		\textbf{Decision}: Benign \\ \hline
		\multicolumn{1}{c}{\textbf{scenario: Data leakage}} \\ \hline
		\textbf{Logon}: Logs in 4 times, above average.\\
		\textbf{Website}: Visited untrusted website "http://wikileaks.org/xxx.php".\\
		\textbf{Drive}: Drive usage within normal range.\\
		\textbf{Suspicious}: Visiting "http://wikileaks.org/xxx.php" is suspicious.\\
		\textbf{Basis of judgment}: Legitimacy of website visits.\\
		\textbf{Decision}: Malicious \\ \hline
		\multicolumn{1}{c}{\textbf{scenario: Insider keylogging}} \\ \hline
		\textbf{Logon}: Logs in 3 times, above average.\\
		\textbf{Website}: Visited websites related to keylogging and downloaded suspicious payloads.\\
		\textbf{Device}: The user has a device usage frequency of 2.0 per day, which is within the average range.\\
		\textbf{Email}: The content expresses dissatisfaction, claiming it's irreplaceability.\\
		\textbf{Download}: Executable file downloaded.\\
		\textbf{Suspicious}: The high logon frequency, the content of the email, and the executable file downloaded.\\
		\textbf{Basis of judgment}: The combination of logon, job dissatisfaction, and download of executable file.\\
		\textbf{Decision}: Malicious \\ \bottomrule
	\end{tabularx}
%	\vspace{5pt}
\end{table}

Improving explanations for the model's detection of insider threats has been challenging because deep learning methods are often viewed as black boxes with opaque decision-making processes. 
This complicates the task for log auditors who need to quickly identify and prevent insider threats.
To address \textbf{RQ4}, we present responses generated by Audit-LLM in three scenarios of insider threats from the CERT dataset, specifically the final responses generated by the Executor. 
These three scenarios are concurrently present in both CERT r4.2 and r5.2 datasets, namely, benign, data leakage and insider keylogging.
Note that due to space constraints, we have made some simplifications to the response generated by Audit-LLM, removing redundant words while preserving the original meaning.
The results are presented in Table~\ref{response}.

As shown in Table~\ref{response}, Audit-LLM is capable of providing accurate responses for various subtasks and accurately identifying decisive factors based on the outcomes of these sub-tasks.
For example, in the scenario of insider keylogging, Audit-LLM identified three results: above-normal login frequency, abnormal email content, and suspicious executable file downloads. 
It combines these findings to conclude that the user exhibited anomalous behavior.
Relatively, when there is no email correspondence or file downloads recorded in the logs, Audit-LLM refrains from conducting corresponding checks, thereby reducing computational overhead.
Further, when all sub-tasks do not exhibit anomaly results, Audit-LLM arrives at a benign conclusion.
In summary, Audit-LLM can provide auditors with audit opinions that are comprehensible to humans. 
Research has shown that automated auditing methods based on LLM can offer certain remediation suggestions~\citep{DBLP:conf/hpcc/QiHLYFYQSXW23}. 
However, in our pilot experiments, we find that without the assistance of a specialized knowledge base, these suggestions are often not applicable. 
Therefore, we do not discuss them in this section.

\subsection{Case study}
\label{casestudy}

To answer \textbf{RQ5}, we deploy Audit-LLM in an actual operational system environment and conduct simulated penetration tests to evaluate its performance in real-world scenarios.
We gather logs exclusively from the native logging systems of the Linux hosts, mainly authorization logs, HTTP logs, network traffic, and other sources.
In total, we monitored 21 hosts for 5 days which were used daily for product development.
During this period, we simulated four types of penetration test activities including MITRE ATT\&CK framework's \textbf{Forge Web Credentials} (T1606), \textbf{Content Injection} (T1586), \textbf{Network sniffing} (T1040), and \textbf{Compromise Accounts} (T1133).
A short description of each attack is shown in Table~\ref{attack}.

Audit-LLM can effectively detect all four types of penetration testing activities. 
Specifically, for \textbf{Compromise Accounts}, Audit-LLM identifies numerous failed login attempts within a short timeframe in authorization logs, triggered by testers attempting brute-force login attacks on target hosts. 
For \textbf{Content Injection}, Audit-LLM performs a thorough analysis of HTTP logs to detect content injection attacks. 
It examines both the payload transmitted in requests and the content retrieved in responses, pinpointing suspicious and potentially malicious activities indicative of content injection attempts. 
Furthermore, for \textbf{Forge Web Credentials}, Audit-LLM ensures comprehensive detection capabilities by meticulously analyzing header information, cookies, session tokens, and authentication parameters to identify Forge web credentials attacks.
Moreover, for \textbf{Network sniffing}, Audit-LLM can identify explicit network sniffing behaviors by analyzing network traffic logs. 
For example, this occurs when a host sends numerous requests using different protocols to various target hosts but fails to receive responses for all of them.

\begin{table}[t]
	\small
	\centering
	\caption{Real-world attack scenarios with short descriptions.}
	\label{attack}
	\begin{spacing}{0.85}
		\begin{tabularx}{\columnwidth}{llX}
			\hline
			\multicolumn{1}{c}{Technique} & \multicolumn{1}{c}{ID} & \multicolumn{1}{c}{Description}                                                                                             \\ \hline
			\begin{tabular}[c]{@{}l@{}}\\Compromise \\ accounts\end{tabular}                 & T1586                  & Adversaries infiltrate existing accounts through various means, such as brute-force attacks, to obtain account credentials. \\
			\begin{tabular}[c]{@{}l@{}}\\Content \\ injection\end{tabular}                   & T1659                  & Adversaries inject malicious content into network traffic to gain system access and communicate with victims.               \\
			\begin{tabular}[c]{@{}l@{}}\\Forge web \\ credentials\end{tabular}               & T1606                  & Adversaries may forge web cookies that can be used to gain access to web applications or Internet services.                 \\
			\begin{tabular}[c]{@{}l@{}}\\Network \\ sniffing\end{tabular}               & T1040                  & Network sniffing refers to using the network interface on a system to monitor or capture information sent over a wired or wireless connection.                 \\ \hline
		\end{tabularx}
	\end{spacing}
%	\vspace{5pt}
\end{table}

\subsection{Cost analysis}

\begin{figure}
	\includegraphics[width=0.49\textwidth]{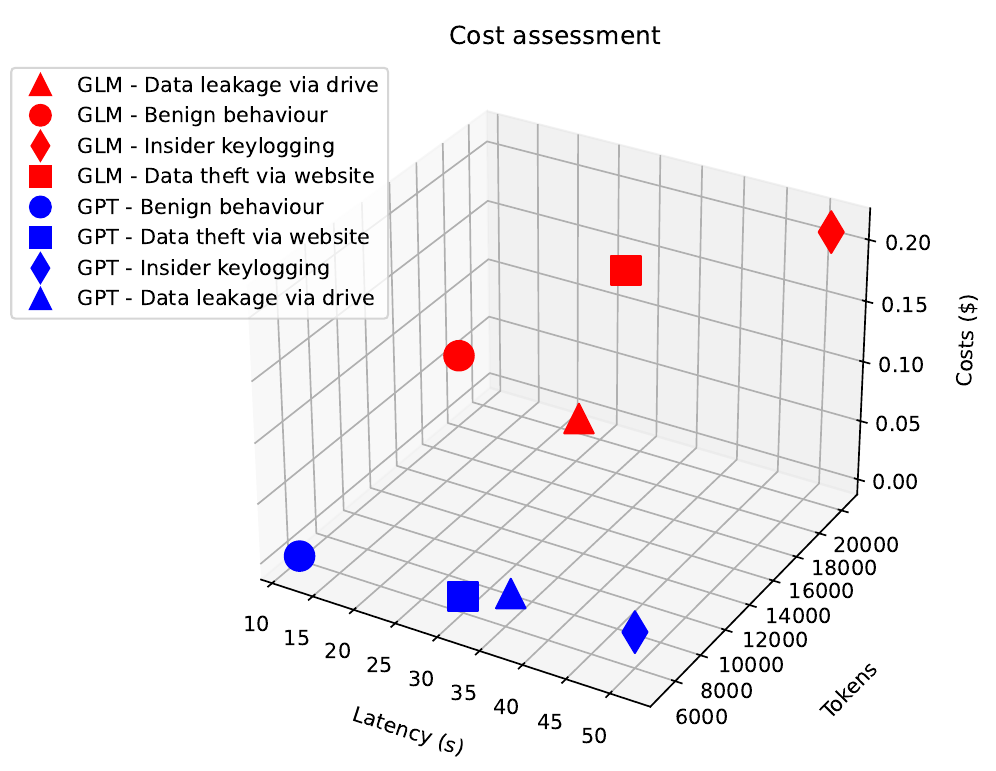}
	\caption{The average latency, token usage, and economic costs of two online LLM APIs across four scenarios.}
	\label{cost}
\end{figure}

Thanks to online APIs, researchers can employ LLMs with extensive parameters even without having access to substantial GPU memory. 
However, the time and economic costs incurred by invoking these APIs remain a significant concern.
To answer \textbf{RQ6}, we present in Figure~\ref{cost} the average latency, token usage, and economic cost of the two online LLM APIs used in our experiment: GPT-3.5 from  OpenAI and GLM-4 from ZhipuAI, across four scenarios, namely, Benign behavior, Data leakage via drive, Data theft via Website, and Insider keylogging.
All data in the experiment was recorded by Langsmith~\citep{langsmith}.

Overall, the cost of using LLM APIs for log analysis is substantial. 
Specifically, in the Insider Keylogging scenario, GPT-3.5 and GLM-4 consumed an average of 6,903 and 20,313 tokens per single CoT process, respectively, costing 0.004\$ and 0.21\$ per transaction.
Comparatively, Insider keylogging and Data theft via website scenarios incur the highest token usage. 
This is because internal threats often involve extensive web browsing, necessitating Audit-LLM to validate the legitimacy of each website accessed.
From the perspective of the base LLM, GPT-3.5 consumes fewer tokens compared to GLM-4. 
This is because the Langchain framework is more compatible with GPT-3.5, providing readily available interfaces for direct invocation. 
In contrast, the adaptation for GLM-4 is less developed, requiring indirect tool invocation through JSON formats.

\label{costanalysis}

%!TEX root = ./Audit-LLM.tex
\section{Conclusion and future work}

In this study, we have proposed a multi-agent log-based insider threat detection framework called Audit-LLM.
Audit-LLM consists of three agents: the Decomposer, the Tool Builder, and the Executor. 
The Decomposer agent breaks down ITD into sub-tasks based on user activity types to construct a Chain-of-Thought (CoT) for auditing.
The Tool Builder agent creates a set of task-specific, reusable tools for extracting contextual information beyond the input window.
The Executor agent uses the generated tools to complete sub-tasks, classify user behavior, and provide human-readable justifications.
Based on experimental results, Audit-LLM effectively leverages the inherent knowledge repository and natural language comprehension abilities of LLM to tackle log-based insider threat detection issues, while avoiding overfitting problems caused by class imbalance. 
Additionally, tools generated by the Tool Builder can efficiently extract statistical log information beyond the input window, thereby assisting the Executor in decision-making.
Furthermore, to address the faithfulness hallucination issue, we propose pair-wise Evidence-based multi-agent debate. 
This approach involves two independent Executors to exchange results obtained from sub-tasks, and iteratively refine their results to improve outcome fidelity.

In our future work, we aim to refine agent design to enhance the accuracy of internal threat detection, minimize false positives, and reduce token usage. Additionally, we intend to integrate a network threat knowledge base akin to MITRE ATT\&CK~\citep{mitre}. 
Our goal is to explore the application of Retrieval Augmented Generation (RAG) to provide network security auditors with effective mitigation recommendations for addressing insider threats.

\bibliographystyle{IEEEtranN}
\bibliography{IEEEabrv,Audit-LLM}

\end{document}